\documentclass{emulateapj}			

\slugcomment{Submitted to AJ}

\usepackage{natbib}
\usepackage[caption=false]{subfig}   

\shorttitle{AO Imaging of VY CMa with LBT/LMIRCam}

\begin{document}

\title{Adaptive Optics Imaging of VY Canis Majoris at 2 - 5 micron with LBT*/LMIRCam}

\altaffiltext{*}{The LBT is an international collaboration among institutions in the United States, Italy and Germany. LBT Corporation partners are: The University of Arizona on behalf of the Arizona university system; Istituto Nazionale di Astrofisica, Italy; LBT Beteiligungsgesellschaft, Germany, representing the Max-Planck Society, the Astrophysical Institute Potsdam, and Heidelberg University; The Ohio State University, and The Research Corporation, on behalf of The University of Notre Dame, University of Minnesota and University of Virginia.}

\author{Dinesh P. Shenoy\altaffilmark{1}, Terry J. Jones\altaffilmark{1}, Roberta M. Humphreys\altaffilmark{1}, Massimo Marengo\altaffilmark{2}, Jarron M. Leisenring\altaffilmark{3}, Matthew J. Nelson\altaffilmark{4}, John C. Wilson\altaffilmark{4}, Michael F. Skrutskie\altaffilmark{4}, Philip M. Hinz\altaffilmark{5}, William F. Hoffmann\altaffilmark{5}, Vanessa Bailey\altaffilmark{5}, Andrew Skemer\altaffilmark{5}, Timothy Rodigas\altaffilmark{5}, Vidhya Vaitheeswaran\altaffilmark{5}}
\affil{$^{1}$ Minnesota Institute for Astrophysics, University of Minnesota, 116 Church St. SE, Minneapolis, MN 55455, USA \\ email: shenoy@astro.umn.edu }
\affil{$^{2}$ Department of Physics, Iowa State University, Ames, IA 50011, USA}
\affil{$^{3}$ Institute for Astronomy, ETH, Wolfgang-Pauli-Strasse 27, 8093 Zurich, Switzerland}
\affil{$^{4}$ Department of Astronomy, University of Virginia, 530 McCormick Road, Charlottesville, VA 22904}
\affil{$^{5}$ Steward Observatory, University of Arizona, 933 N. Cherry Ave, Tucson AZ 85721, USA}

\newcommand{\Ks}{K$_s$}
\newcommand{\Lp}{L$'$}

\begin{abstract}
We present adaptive optics images of the extreme red supergiant VY Canis Majoris in the \Ks, \Lp ~and M bands (2.15 to 4.8 $\micron$) made with LMIRCam on the Large Binocular Telescope (LBT).  The peculiar ``Southwest Clump'' previously imaged from 1 to 2.2 $\micron$ appears prominently in all three filters.  We find its brightness is due almost entirely to scattering, with the contribution of thermal emission limited to at most $25$\%.  We model its brightness as optically thick scattering from silicate dust grains using typical size distributions.  We find a lower limit mass for this single feature of 5 $\times$ $10^{-3}$ $M_{\sun}$ to 2.5 $\times$ $10 ^{-2}$ $M_{\sun}$ depending on the assumed gas-to-dust ratio.  The presence of the Clump as a distinct feature with no apparent counterpart on the other side of the star is suggestive of an ejection event from a localized region of the star and is consistent with VY CMa's history of asymmetric high mass loss events. \\
\end{abstract}


\section{INTRODUCTION}
The extreme red supergiant VY Canis Majoris is one of the brightest 5 - 20 $\micron$ stellar sources in the sky.  Recent observations of SiO masers in its circumstellar envelope place it at a distance of 1.2 kpc using trigonometric parallax \citep{Zhang:2012}, with $L_{bol} = 2.7 \times 10^{5} L_{\sun}$ \citep{Wittkowski:2012}.   Space-based observations with the \emph{Hubble Space Telescope (HST)} from 0.4 to 1 $\micron$ have catalogued an extensive, highly structured nebula consisting of multiple arcs and knots of material ejected within the past $\sim$1000 years.  This asymmetric nebula is indicative of multiple episodes of localized mass ejections from active regions on its surface \citep{Smith:2001, Humphreys:2005, Humphreys:2007}.  Ground-based adaptive optics imaging with the Large Binocular Telescope (LBT) enables us to extend the exploration of VY CMa's mass loss into the near-IR at an angular resolution greater than previous studies, e.g. \cite{Cruzalebes:1998, Monnier:1999}.  

\begin{figure}[ht]
\centering
\subfloat[]{\label{Cruzalebes}\includegraphics[scale=0.35]{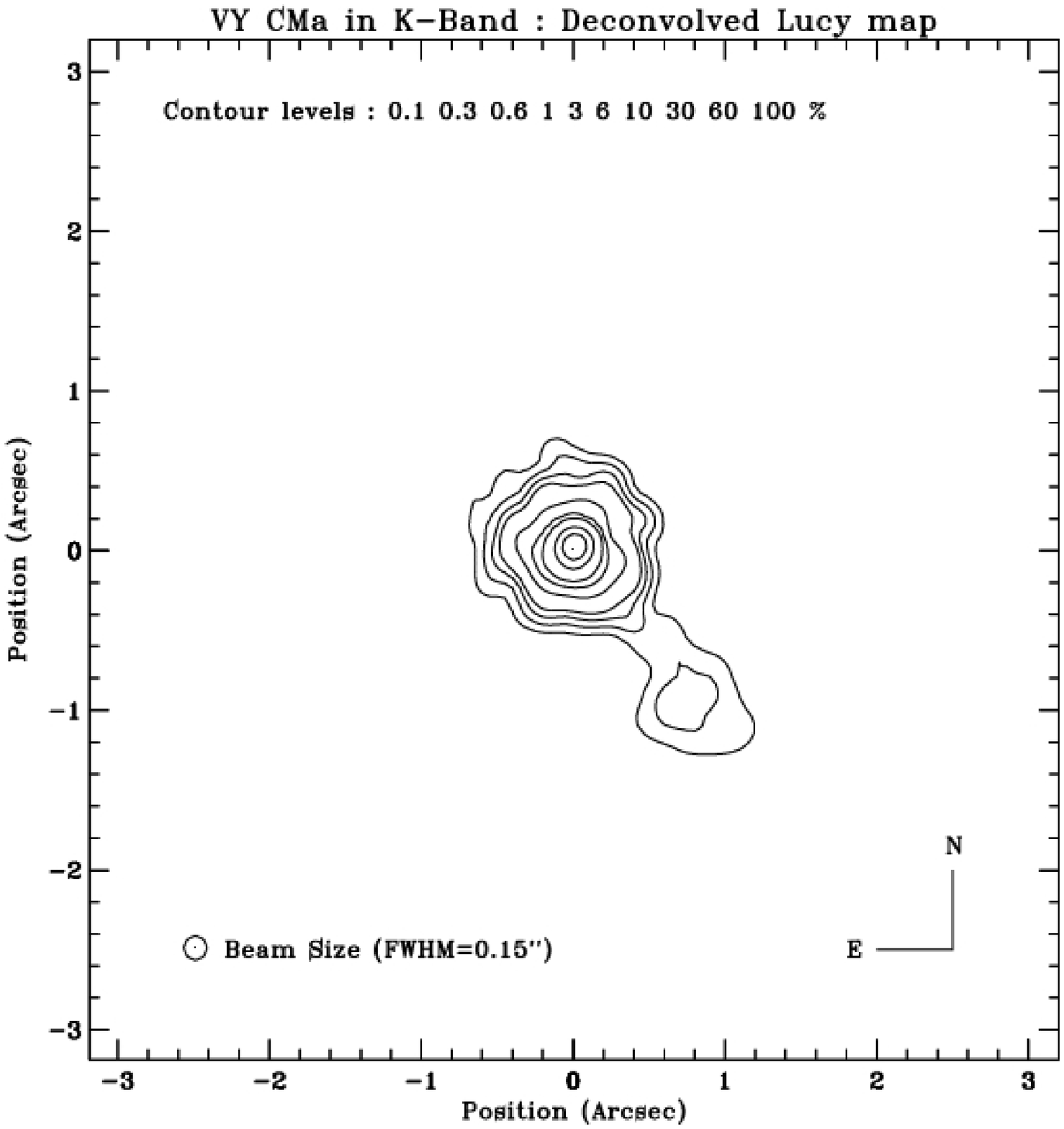}}
\\
\subfloat[]{\label{3DMorphIFig4}\includegraphics[scale=0.35]{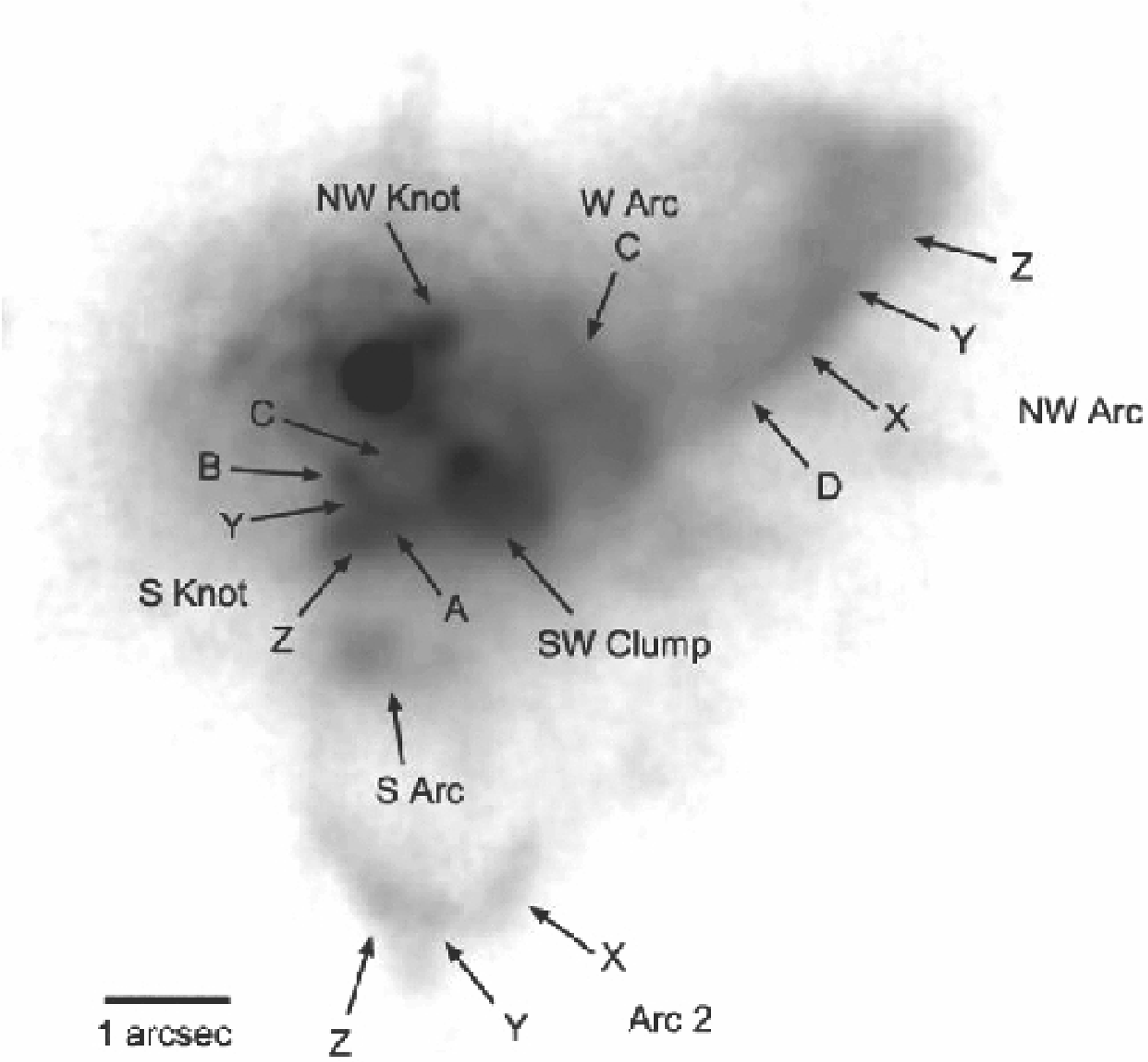}}
\caption{\emph{Top:} K-band image of VY CMa from Figure 3 of \citet{Cruzalebes:1998}, taken in 1994 with the COME-ON+ AO system at the ESO-La Silla 3.6m telescope.  The authors noted a ``quasi-circular clump'' located $\sim$ 1$\arcsec$ to the southwest of the star.  \emph{Bottom:}  2005 \emph{HST} F1042M (1 $\micron$) image of VY CMa from Figure 4 of \citet{Humphreys:2007}, identifying the SW Clump as well as numerous other features of the ejecta.  Several of the features at larger separations from the star than the SW Clump are potentially observable with LMIRCam. }
\end{figure}

Adaptive optics images of VY CMa made in 1994 with the COME-ON+ AO system at the ESO-La Silla 3.6m telescope found  a ``quasi-circular clump'' of emission in the K-band at approximately $1\arcsec$ to the southwest of the star \citep{Cruzalebes:1998}(see Figure \ref{Cruzalebes}).  J-band (1.25 $\micron$) imaging in 1996-97 with the ADONIS SHARPII+ camera showed a similarly bright knot in the same position \citep{Monnier:1999}.  The clump of material was also observed at 1 $\micron$ in \emph{HST} imaging in 1999 \citep{Smith:2001} (hereafter S01), and again in 2005 \citep{Humphreys:2007} (hereafter H07) as part of multi-epoch observations to map the morphology of VY CMa's nebula.  S01 \& H07 refer to it as the Southwest (SW) Clump (see Figure \ref{3DMorphIFig4}).    

\citet{Monnier:1999} suggested the SW Clump might be component ``B'' that had been reported as a possible moving companion star by early observers.  However, S01 noted that the SW Clump and another clump directly South were equally close to the position of ``B'' as reported by \citet{Herbig:1972}.  S01 could not unambiguously identify any of the structures in the HST images with the visual ``companions'' in the early observations. They concluded that the condensations or knots were either changing their structure and positions in only a few decades or that variable illumination was altering the appearance of the clumps \citep{Wallerstein:1978}.

Combining slit spectroscopy from \citet{Humphreys:2005} with the \emph{HST} images across two epochs, H07 computed radial and transverse velocities of numerous components of the ejecta, including the SW Clump.  \citet{Jones:2007} used imaging polarimetry from the $HST$ observations to locate the physical positions of several of the features in the ejecta including the SW Clump, and independently confirmed the three-dimensional geometry of VY CMa's nebula.  There is some ambiguity as to whether the SW Clump is a single feature or  a superposed image of two distinct parts of the ejecta moving away from the star along different radial trajectories.  H07 observed several knots of ejecta in the \emph{HST} visual images at the same location as the SW Clump in the $HST$ 1 $\micron$ images and determined them to be moving towards us.  In their analysis H07 reasoned that since the SW Clump is highly obscured (appearing only at 1 $\micron$, not in any of the \emph{HST} WFPC2 visual filters), the more redshifted velocities from the slit spectroscopy applied to it.  For VY CMa's distance of 1.2 kpc, H07's observations place the SW Clump on a trajectory inclined +10$\degr$ behind the plane of the sky, moving at $\sim$ 14 km/s after having been ejected about 500 years ago.  

\section{OBSERVATIONS AND REDUCTION}
\subsection{Observations}
We observed VY CMa on UT 2011 November 16 with LMIRCam \citep{Skrutskie:2010} using a single 8.4 m primary mirror on the Large Binocular Telescope (LBT).  At the time of these observations LMIRCam used an HgCdTe array of 1024 $\times$ 1024 pixels, with a pixel scale of 0.011$''$ pix$^{-1}$ for a field of view $\thicksim$ $11'' \times 11''$ \citep{Leisenring:2012}.  Images were made in the \Ks ~($\lambda_0$ = 2.15 $\micron$), \Lp ~($\lambda_0$ = 3.8 $\micron$) and M ($\lambda_0$ = 4.8 $\micron$) filters.

To assess the point spread function and the orientation of the images, the binary star HD 37013 was observed in the \Lp ~filter.  The PSF exhibited astigmatism in the form of a ``cross'' pattern, which has been attributed to poor mounting of the dichroic \citep{Rodigas:2012}.  The LBT's first-light adaptive optics (AO) system \citep{Esposito:2012} provided near-diffraction-limited imaging, with the first minimum in the Airy disk pattern of the primary star in HD 37013 at \Lp ~occuring at $0\farcs14$ (compared to the theoretical minimum of $0\farcs11$ computed from $\theta$ = 1.22$\lambda/D$).

We imaged VY CMa at four dither positions on the array.  Images at each dither position were taken in sets of 10 frames with 2.9 seconds per frame.  The 2.9 second exposure time allowed fainter regions at greater than about $0\farcs5$ from the star to be imaged at the expense of saturating closer to the star.  Shorter length calibration exposures of 0.29 seconds were taken through a neutral density (ND) filter which attentuates the transmitted light by a factor of 100.  These exposures provided unsaturated images of the star to use for flux calibrating ADUs, as well as for alignment of the saturated images.  Nearby sky images were taken for both sets of exposure times for background subtraction.  

\subsection{Reduction}
We examined the individual sets of ten frames obtained in each dither position, discarding frames in which it appeared that the AO lock had broken during an exposure.  The remaining images were coadded to make a single image of VY CMa in each of the four dither positions.  The nearby sky images were similarly coadded.  Bad pixels were masked by mean combining dark images and masking pixels whose value deviated by $>3\sigma$ from the median in the combined dark image.  Masked pixels were replaced with the median value of the 8 surrounding pixels.  We removed most of the sky background and offset level by subtracting the nearby sky images from each VY CMa image, matching filters and exposure times and selecting the sky image which had been taken closest in time.  Subtracting pairs of dither positions to remove the background offset level and hot pixels was not possible because the angular extent of VY CMa's nebula is comparable to LMIRCam's $11'' \times 11''$ field of view.  To bring the background of the four images in each filter (eight images in the M filter) closer to a mean value of zero, we subtracted the average of the sky background counts in four regions around VY CMa (separated from the star by several arcseconds).  

\begin{figure*}[ht]
\centering
\includegraphics[scale=0.5]{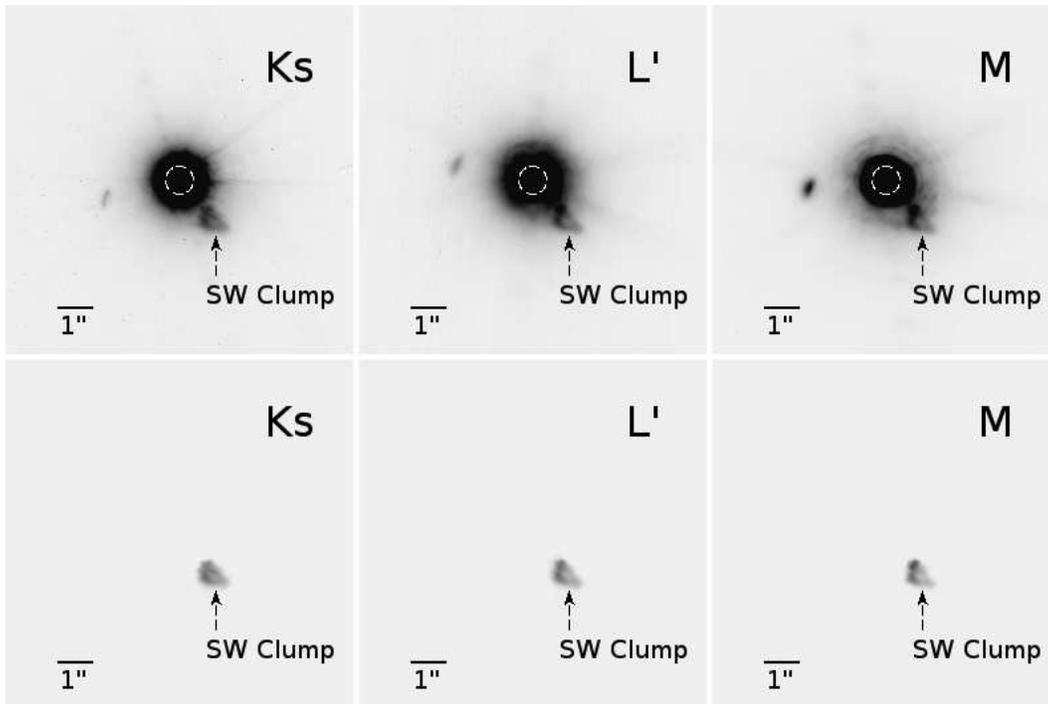} 
\caption{\emph{Top row}:  VY CMa imaged with LMIRCam for 2.9s through the \Ks ~filter ($\lambda_{0}$ = 2.15 $\micron$), \Lp ~filter ($\lambda_{0}$ = 3.8 $\micron$) and M filter ($\lambda_{0}$ = 4.8 $\micron$).  Each image is a $10\arcsec \times 10\arcsec$ FOV oriented North up, East left with linear greyscaling.   The bar in the lower left of each image measures $1\arcsec$ = 1200 AU at a distance of 1.2 kpc.  The beam size at \Lp ~is $0\farcs12$  (measured FWHM of a point source).  The dashed circle in the center indicates where the image is saturated out to a radius of $\sim 0\farcs4$ around the star.  The SW Clump appears bright in all three filters.   The patch of light appearing to the East of the star in each image is presumed to be from to an internal reflection rather than part of VY CMa's ejecta, due to its shifting positions relative to the star between dither positions and between filters.   \emph{Bottom row}: Images of SW Clump after subtracting a projected 2-D profile of the star's flux into the region covering the Clump (see $\S$ 4.1).  \vspace{6pt}}
\label{reduced_ims}
\end{figure*}

LMIRCam was fixed to the telescope during the observations with the rotator turned off, resulting in the field of view rotating in the time between dithers.  Therefore we rotated each dither position's image by the parallactic angle computed for the mean time of observation at that position.  To verify the accuracy of the rotation, we applied the same calculation to the image of binary star HD 37013 that was observed on the same night.  When rotated to take out the parallactic angle so that the image displays with North up and East to the left, the position angle of the SW Clump on the sky in each image agrees with the Clump's position in the $HST$ 1 $\micron$ image. In the absence of any other identifying features in the images, final adjustment of the alignment of the images across the three filters was done by sighting on the distinct triangular shape of the SW Clump itself and assuming it centers on the same position at all three wavelengths.  With the images in each dither position aligned, the four dithered images (eight in the M filter) were coadded.  We show our resulting images in the top row of Figure \ref{reduced_ims}.

\subsection{Flux Calibration}
We used the shorter 0.29s exposures taken through the neutral density (ND) filter to flux calibrate the ADUs (counts) in the 2.9s images.  We obtained background-subtracted counts of the star at \Ks ~from aperture photometry on the ND images with the same 0$\farcs$4 diameter aperture used by S01 on their 2.14 $\micron$ adaptive optics ADONIS/SHARP II+ images of VY CMa made with the ESO 3.6 m telescope.  This aperture diameter equals $\sim4\sigma$ of the star radial profile in the \Ks ~ND image and thus encloses essentially all the flux of the star, as the same diameter did on the ESO images.   The counts in this aperture were scaled to account for the difference in integration time and the attenuation by the ND filter.  We cross-checked the accuracy of this scaling by comparing the average brightness of the SW Clump at M in the 2.9s images with its average brightness in the shorter exposures through the ND filter, where it is marginally visible.  The scaled counts were calibrated to the upper-limit flux in W cm$^{-2}$ at \Ks ~attributed to the star by S01 (the thin solid curve below the total continuum curve at the top of Figure \ref{SED}).   To determine the flux of the star in counts in the \Lp ~and M ND images, we selected aperture diameters to be 4$\sigma$ of the star profile at each of those wavelengths, as was done at \Ks.  This ensured these apertures ($0\farcs49$ at \Lp ~and $0\farcs63$ at M) enclosed the same fraction of total flux as the fraction enclosed at \Ks.  These counts were likewise scaled up and calibrated to the upper-limit fluxes in W cm$^{-2}$ at \Lp ~and M attributed to the star.  Uncertainties in the calibration factors, estimated from the scatter in the total star counts from each of the 3 to 4 dithered images taken through the ND filter at each wavelength, are negligible in comparison to the uncertainty in the flux of the SW Clump as discussed in $\S$ 4.1.

\section{RESULTS}  
\subsection{Comparison to Previous Observations}
The SW Clump appears bright in all three LMIRCam filters, measuring about 1$\arcsec$ across.  Although its appearance overall is generally smooth, there is some variation in brightness across it.  In all three filters the Clump appears somewhat dimmer in its center than at its edges (see Figure \ref{reduced_ims}, bottom; see also the radial profiles in Figure \ref{cut_comp}).   As noted earlier, H07 observed several knots at visible wavelengths which are in the same location as the SW Clump but which are moving towards instead of away from us.  It is possible that the SW Clump as it appears in the LMIRCam filters is a projection of two distinct parts of VY CMa's ejecta which happen to lie along the same line of sight.  On the other hand, J07 found the SW Clump to be a separate feature in their polarimetry map.  In J07's polarimetry map the SW Clump was about 40\% polarized, twice as polarized as the surrounding nebulosity.  This suggested that the SW Clump is a distinct feature.  In our analysis below we adopt the view that the SW Clump is a single feature observed moving along a trajectory inclined +10$\degr$ behind the plane of the sky.

Four other features of VY CMa's ejecta identified in previous observations are located at angular separations from the star greater than that of the SW Clump and are therefore potentially observable with LMIRCam.  These features were labeled by H07 as the NW Arc, the W Arc, the S Arc, and Arc 2 (see Figure \ref{3DMorphIFig4}).  The portions of Arc 2 which H07 designated as X, Y and Z are marginally observed in the LMIRCam \Ks ~image (S/N $\sim$ 2-3).  The portions of the NW Arc designated D, X, and Y are possibly  visible in the \Ks ~filter as well, but are extremely faint.  We do not observe the S Arc and W Arc above the background level in the \Ks ~image.  None of these four features are observed in the \Lp ~or M images.  

Previous ground-based IR images at M, at 8.4 $\micron$, and at 9.7 $\micron$ obtained by S01 with the TIMMI instrument at ESO-La Silla in 1996 appeared to show an axis of extended emission running east-west, with peaks at about 1$\arcsec$ from the star (their Figure 5.)  While noting that these twin emission peaks could be accounted for with thermal emission from dust grains (using VY CMa's then-accepted distance of 1.5 kpc), S01 cautioned that the extended structure could be an artifact of defective PSF subtraction at those wavelengths.  Our higher resolution M-band image with LMIRCam finds no twin peaks of emission present at radii of 1$\arcsec$ to the east and west of the star.

\section{ANALYSIS \& DISCUSSION}
\subsection{Subtraction of Central Star Flux}
To quantify the flux of the SW Clump relative to VY CMa's observed SED we first needed to remove the contribution of the star's flux in the region of the SW Clump.  The proximity of the SW Clump to the star complicates this subtraction.  Furthermore, at \Lp ~the profile of the star's flux is not well represented by the PSF of either of the stars in the HD 37013 binary pair.  The profile of the star's light at \Lp ~is broader in the southeast and southwest than other directions, presumably due in part to emission from unresolved ejecta at less than $1\arcsec$ from the star.  At M the star's profile is similarly extended towards the southwest and would be poorly represented by scaling the PSF at \Lp.

We therefore used two alternate methods for subtracting the star's flux from the Clump, and took the average of the two methods as the Clump's flux for analysis.  While the fluxes of the Clump from the two methods differ by as much as 50\% at M, this difference between the two methods does not change our finding that the Clump's flux at M can be largely accounted for by scattered light alone.

In the first method, we created an azimuthal average of the radial profile of the central region containing the star in each filter.   The range of azimuth covering the SW Clump (position angles from 200$\degr$ to 230$\degr$ E of N) was excluded from this average, as was the range containing the ``ghost'' in the East of the images that is assumed to be due to an internal reflection.  We then rotated the profile of this azimuthal average to make a circularly symmetric image to represent the star's light and subtracted it from the image.    

In the second method, we masked out the region containing the SW Clump  and replaced it with a 2-D surface that approximates the shape of the star's 2-D profile into the region covering the Clump in each filter.  The 2-D star profile surface was made using IDL's TRI\_SURF routine, which uses linear interpolation to create a surface filling in the masked out region.  We subtracted this projected image from the image in each filter.  The subtracted images are displayed in the bottom row of Figure \ref{reduced_ims}.  

\begin{figure}[ht] 
\centering
\includegraphics[scale=0.4]{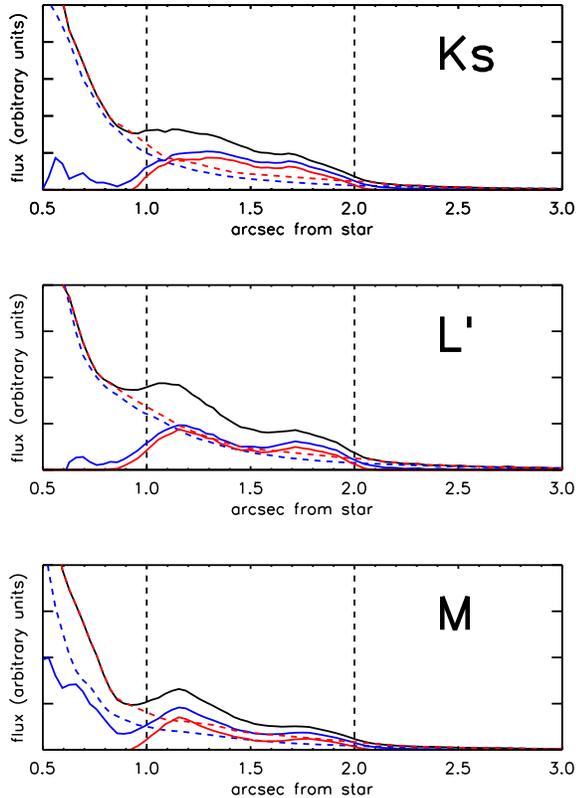}  
\caption{Radial cuts comparing the effectiveness of subtracting the star's flux from the image of the SW Clump in each LMIRCam filter.  The dashed vertical black lines mark the radial range in which the Clump appears.  The solid black curve is the radial profile of the image (star and SW Clump combined).  The dashed blue curve is the profile of an azimuthal average of the combined image.  The solid blue curve is the profile of the combined image minus the azimuthal average image.  The dashed red curve is the profile of a linearly interpolated surface that projects the star's light into the region containing the SW Clump.  The solid red curve is the profile of the combined image minus the 2-D projected image.  Subtracting the 2-D projected image yields an integrated flux of the SW Clump which is $\sim$ 20\% lower at \Ks ~and \Lp, and is $\sim$ 50\% lower at M, as compared to the Clump's flux when subtracting the azimuthal average image. \vspace{6pt} }
\label{cut_comp}
\end{figure}

Figure \ref{cut_comp} compares the two subtraction methods for a cut through the SW Clump.  At each wavelength, subtracting the 2-D projected star profile (red dashed line) results in a lower flux from the Clump than subtracting the azimuthally averaged star profile (blue dashed line).   At \Ks ~and \Lp ~the 2-D projected star profile provides a better match to the likely profile of the star in the region of the Clump, though it may slightly oversubtract at radii close to the star.  The difference in the flux of the SW Clump at \Ks ~and \Lp ~for the two methods is $\sim$ 20\%.  At M the azimuthally averaged star profile is substantially lower than the likely profile of the star in the region of the Clump, which the 2-D projected star profile follows more closely.  The difference in the flux of the SW Clump measured from the two methods is $\sim$ 50\% at M.  

\begin{figure}[ht]  
\centering
\includegraphics[scale=0.33]{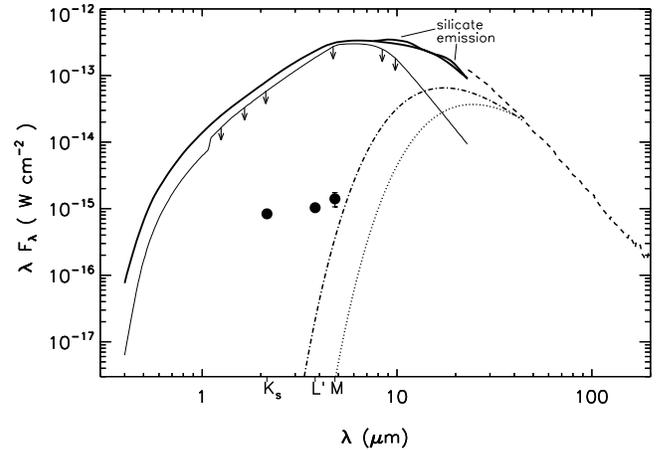}
\caption{Observed flux of SW Clump in LMIRCam filters, plotted on the spectral energy distribution of VY CMa.  The solid thick curve at the top is the total continuum flux of VY CMa (the star and nebula combined) from the $HST$ and ground-based photometry by \citet{Smith:2001} (S01).  The thin solid line below it indicates the flux which S01 attributed to the star/inner dust shell, which are upper limits at IR wavelengths due to limits in spatial resolution.  The discontinuity in the star's flux at $\sim$ 1 $\micron$ resulted from the different aperture sizes used by S01 to measure the flux of the star, corresponding to the difference in resolution between their $HST$ and ground-based observations.  The dashed line from 23 $\micron$ to 200 $\micron$ is a portion of the ISO spectrum of VY CMa from \citet{Harwit:2001}.  The discontinuity in VY CMa's continuum flux at 23 $\micron$ (a factor of $\sim$ 1.4) may be due ISO's larger beam size.   The solid circles are the flux of the SW Clump in the LMIRCam \Ks, \Lp, and M filters (this work).  The uncertainty in its flux in each filter is taken to be the difference of the two methods of star flux subtraction discussed in $\S$ 4.1.  The uncertainty in its flux at \Ks ~and \Lp ~is smaller than the plotted circle.  The dotted curve is a $\lambda B_{\lambda}(T_{eq})$ thermal emission spectrum for grains at $T_{eq}$ = 150 K, the temperature we adopt for grains in the SW Clump.  The dot-dash curve is for an upper limit temperature of $T_{eq}$ = 210 K.  The thermal emission curves are plotted out to $\lambda$ = 45 $\micron$, the wavelength through which VY CMa's far-IR spectrum constrains the contribution of thermal emission to the Clump's observed flux (see $\S$ 4.2).  For even the upper limit temperature of $T_{eq}$ = 210 K, three-quarters of the Clump's flux at M must be due to non-thermal emission.  \vspace{6pt}}
\label{SED}
\end{figure}

While these differences may be substantial, they do not affect our conclusion that the Clump's brightness can be accounted for largely with scattered light alone.  For the following analysis of the SW Clump we adopt the average flux from the two star subtraction methods.  In Figure \ref{SED} we plot the flux of the SW Clump from the average of the two subtraction methods on VY CMa's SED.   As a check on our calculation, we compared our flux values with those obtained by S01 from their lower resolution ADONIS/SHARP II+ ESO image of VY CMa at 2.14 $\micron$.  Our flux-calibrated average brightness of the SW Clump from the LMIRcam \Ks ~image agrees to within a factor of 2-3 with the surface brightness at the matching location of the SW Clump in the ESO images. 
	
\subsection{Constraint on Thermal Flux of the SW Clump}
LMIRCam's \Ks, \Lp ~and M filters sample the wavelength range where emission from dust grains may be due to either scattering, thermal emission, or a combination of both.  Over most of the likely range of the Clump's temperature, we find that thermal emission contributes negligibly to its flux in all three filters. This conclusion is driven by two factors: the very bright fluxes at \Ks ~and \Lp ~which are due to scattered light, and the far-IR flux of the entire system, which constrains the thermal contribution from the Clump.

We determine an upper limit on the grains' thermal flux by assuming them to be silicates and computing their equilibrium temperature $T_{eq}$ for their distance $d$ from the star.  In computing $T_{eq}$, we include the grains' wavelength dependent absorption efficiency $Q_{abs}(\lambda)$ when balancing the power from the star absorbed by a grain with the power emitted by the grain.  We use $Q_{abs}(\lambda)$ for silicate grains of mean radius $\overline{a}$ = 0.3 $\micron$, which is a typical size for circumstellar environments (see $\S$ 4.3).  Parts of the Clump could be at distances varying as much as from 1200 AU to 2400 AU from the star, corresponding to equilibrium temperatures from 170 K down to 130 K.  A distance measured to the middle of the face of the Clump puts it at roughly $d$ = $1\farcs4$ = 1600 AU from the star.  For this average distance the grains' equilibrium temperature is $T_{eq}$ = 150 K.  This is slightly higher than the blackbody equilibrium temperature of 140 K at $d$ = 1600 AU since $Q_{abs}(\lambda) < 1$ makes the grains less-than-ideal radiators.  Including $Q_{abs}(\lambda)$ does not raise their equilibrium temperature much, however.  If the grains were subjected to the flux of a star whose SED peaked in the visible where $Q_{abs}(\lambda)$ is closer to 1, they might reach a higher equilibrium temperature than 150 K.   Varying $\overline{a}$ over a reasonable range changes $Q_{abs}$ somewhat, but the resulting $T_{eq}$ is essentially the same. 

Given the complexity of VY CMa's circumstellar environment, the temperature of the dust grains may lie within a wider range than the 130 K to 170 K obtained from simply placing grains in clear sight of the star and having them freely re-emitting into space.  In the 8.4 and 9.8 $\micron$ images obtained by S01 the respective brightnesses at the location of the SW Clump  yield a color temperature of 210 K (their Figure 5).   In view of our finding further below that the fluxes at \Ks ~and \Lp ~are dominated by scattered light and given that the star's SED continues to rise from \Ks ~ through $\sim$ 8 $\micron$, the infrared color at 8.4 $\micron$ must include a significant scattered light component.  If the 8.4 $\micron$ brightness is partly due to scattered light, that would lower the $S_{8.4}/S_{9.8}$  color temperature at the Clump's location to below 210 K, though it might still be higher than 150 K.

The grains could be cooler than $T_{eq}$ = 150 K as well.  Recent millimeter wavelength observations of VY CMa's molecular emission have been used to derive gas temperatures which, at the SW Clump's distance from the star, range from 150 K down to as low as 80 K \citep{Muller:2007, Fu:2012}.   Grains with temperatures as cold as 80 K cannot contribute to the Clump's observed flux in the M filter because the extrapolated flux at longer wavelengths would far exceed what is observed for the entire system. For completeness we consider the impact of a grain temperature between 80 K to 210 K on the Clump's observed flux at M.

The contribution of thermal flux to the Clump's brightness in the M filter is constrained by the grains' temperature as well as by the wavelength at which the Clump becomes optically thin in emission.  For temperatures from 80 K to 210 K the Clump's $B_{\lambda}(T_{eq})$ thermal spectrum rises through the M filter to a peak at a longer wavelength.  This thermal spectrum cannot exceed VY CMa's observed far-IR spectrum, which was measured out to 200 $\micron$ by ISO \citep{Harwit:2001}.    In the next subsection we find that the Clump is optically thick at \Ks, \Lp ~and M.  If the Clump becomes optically thinner in emission at longer wavelengths, the spectrum of its thermal emission would be reduced to $(1 - e^{-\tau_{abs}(\lambda)})\cdot B_{\lambda}(T_{eq})$ at those longer wavelengths.  Depending on the wavelength at which this transition occurs, this permits a higher thermal flux at M without exceeding VY CMa's far-IR SED.  

To keep it simple, consider the transition to occur when the absorption optical depth $\tau_{abs}$ drops by a factor of 2, with the scaling given by:  $\tau_{abs}(\lambda) = \tau_{abs}(M)\cdot  Q_{abs}(\lambda) / Q_{abs}(M)$.  For our grain parameters we find this transition occurs at $\lambda \approx$ 45 $\micron$.  In Figure \ref{SED} we plot two $\lambda B_{\lambda}(T_{eq})$ curves for thermal emission that is optically thick out to this wavelength.    One curve is for our adopted value of $T_{eq}$ = 150 K (dotted line) and the other is for the upper end $T_{eq}$ = 210 K (dot-dashed line).  Ignoring any contribution from the rest of VY CMa's nebula to the flux at 45 $\micron$, these curves constrain the maximum thermal flux from the Clump at M for each of these temperatures.  For $T_{eq}$ = 150 K the thermal flux at M would be only 0.2\% of the Clump's observed flux at M.  For the upper end $T_{eq}$ = 210 K the thermal flux at M would optimistically be 25\% of the Clump's observed flux at M.  Thus even if the Clump's total extinction optical depth at M is no greater than order unity, three-quarters of its flux must be attributed to non-thermal emission, i.e., to scattered light.

\subsection{Scattered Light Flux of the SW Clump}   
In modeling the brightness of the Clump as due to scattering, we first consider the possibility that the Clump is optically thin at \Ks, \Lp ~and/or M.  Surface brightness at a given wavelength due to optically thin scattering may be computed with 
\begin{equation} 
S_\lambda = \frac{ \omega \tau_{ext} F^\star_\lambda \sin^2\theta }{ 4\pi \phi^2 } \Phi(\theta,\lambda)
\end{equation} 
where $\omega \tau_{ext} = \tau_{sc}$ is the scattering optical depth at $\lambda$, $F^\star_\lambda$ is the flux of the star at $\lambda$, $\theta$ is the scattering angle, $\Phi(\theta,\lambda)$ is a wavelength-dependent phase function accounting for the variation with $\theta$ of the intensity of scattered light from silicate spheres, and $\phi$ is the angular distance between the star and a position on the Clump \citep{Sellgren:1992}.  Taking $\tau_{sc} \lesssim$  0.2 as a reasonable upper limit for being optically thin, we find the brightness across the SW Clump is an order of magnitude brighter at \Ks ~than can be explained by optically thin scattering.  At \Lp ~and M as well the Clump is brighter than optically thin scattering of the star's light can account for, even at the Clump's upper limit temperature of $T_{eq}$ = 210 K for which thermal emission would supply a quarter of its brightness at M.

We adopt the view that the SW Clump is moving away from the star and behind the plane of the sky.  We may therefore treat it as a plane parallel atmosphere whose surface brightness arises from optically thick scattering (diffuse reflection).  Surface brightness at a given wavelength due to diffuse reflection from a plane parallel atmosphere may be expressed as: 
\begin{equation} 
S_\lambda = \frac{ \overline{\omega_0}(\lambda) F^\star_\lambda \sin^2\theta }{ 4\pi \phi^2 }  \left(\frac{\mu_0}{\mu+\mu_0}\right) H(\mu)H(\mu_0)
\end{equation}
where $\overline{\omega_0}(\lambda)$ is the particles' single scattering albedo at $\lambda$, $F^\star_\lambda$, $\theta$ and $\phi$ are the same as in the previous equation, $\mu_0$ and $\mu$ are respectively the direction cosines of the incident and reflected beams measured from the normal to the plane parallel atmosphere, and the $H$ terms are numerically integrated functions whose value range from 1 up to $\sim$3 for increasing $\mu$ and albedo \citep{Chandrasekhar:1950}.  The angles of the incident and reflected beams of light in/out of the Clump are assumed to be equal ($\mu_0$ = $\mu$ = $\cos 40\degr$, based on the inclination of the Clump's trajectory at +10$\degr$ behind the plane of the sky.)    

We idealize the grains to be silicate spheres of radius $a$, with the distribution of grain radii following either a power law distribution $n(a)da \propto a^{-p}da$, or a Gaussian distribution with mean radius $\overline{a}$ and standard deviation $\sigma_{a}$.  Using the BHMIE subroutine\footnote{\footnotesize{http://www.met.tamu.edu/class/atmo689-lc/bhmie.pro}} of \citet{BohrenHuffman:1983}, we compute the grains' extinction and scattering efficiencies numerically from Mie theory using optical constants $n(\lambda)$ and $k(\lambda)$ taken from the literature, and then compute the albedo with $\overline{\omega_{0}}(\lambda) = Q_{sc}(\lambda) / Q_{ext}(\lambda)$.  The optical constants $n$ and $k$ are the wavelength-dependent real and imaginary parts of silicates' complex index of refraction $m(\lambda) = n + ik$.  With the computed albedo $\overline{\omega_0}(\lambda)$ for a chosen power law or Gaussian distribution we compute a model SED of the SW Clump at each wavelength using Equation (2).  This model SED is then scaled to fit the observed flux of the Clump at either \Ks ~and/or M.

\begin{figure}
\centering
\includegraphics[scale=0.4]{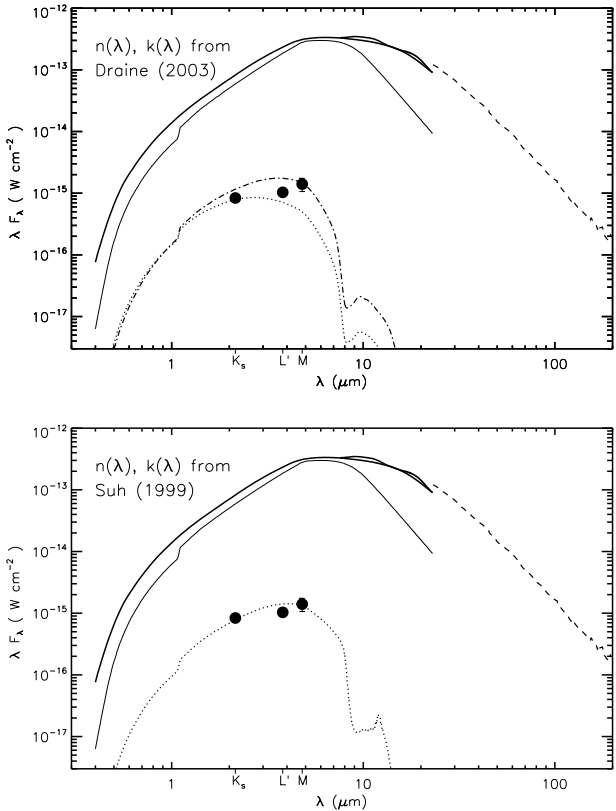}
\caption{Examples of modeling SW Clump as diffuse reflection of the star's light from silicate grains with a distribution of radii.  \emph{Top:} Models using ``astronomical silicate'' optical constants $n(\lambda), k(\lambda)$ from \citet{Draine:2003a} for power law distributions with index $p$ = $-3.0$.  Minimum and maximum radii for the distributions have been chosen to fit the Clump's \Ks ~flux (dotted curve) and M flux (dash-dotted curve).  For these $n$, $k$ no distribution was found which fit the flux at both \Ks ~and M.  \emph{Bottom:} Model using $n(\lambda)$ and $k(\lambda)$ from \citet{Suh:1999} for silicate grains in the dust shells of AGB stars.  A power law distribution of grain radii with index $p$ = $-3.0$ over the range $a_{min}$ = 0.2 $\micron$ to $a_{max}$ = 0.5 $\micron$ is able to fit the Clump's at both \Ks ~and M (dotted curve).  For both sets of optical constants, Gaussian distributions of the grain radii with mean radius $\overline{a}$ $\approx$ 0.3 $\micron$ are able to produce essentially the same results. \vspace{6pt}} 
\label{DR_stack}
\end{figure}

In Figure \ref{DR_stack} we have plotted examples of these diffuse reflection SEDs fitted to the Clump's observed flux at either \Ks ~and/or M for selected power law and Gaussian distributions of the grain radii.  Using optical constants $n(\lambda)$ and $k(\lambda)$ for ``astronomical'' silicates from \citet{Draine:2003a} we can fit the flux at \Ks ~or at M with power law and Gaussian size distributions, but not at both wavelengths (Figure \ref{DR_stack}, top).  We obtain a better fit when using the optical constants $n(\lambda)$ and $k(\lambda)$ computed by \citet{Suh:1999} for silicate grains in the dusty envelopes around AGB stars.  With these constants we can fit the Clump's flux at both \Ks ~and M using a single power law or single Gaussian size distribution, for both of which $\overline{a}$ $\approx$ 0.3 $\micron$ (Figure \ref{DR_stack}, bottom).  For the high-end $T_{eq}$ = 210 K at which thermal flux would supply a quarter of the total observed flux at M, a slightly tighter grain size distribution fits the lowered diffuse reflection flux at M.  The difference in size distributions is minor and the mean radius remains $\overline{a}$ $\approx$ 0.3 $\micron$.  In all cases fitting a diffuse reflection SED to the flux at M within the range of its uncertainty results in overpredicting the flux at \Lp.  Nonetheless, the reasonably good agreement between the predicted diffuse reflection fluxes for these typical grain sizes and the observed flux at \Ks ~and M indicates that the Clump's brightness in all three filters is due largely or solely to scattered light.
 
\vspace{6pt}
\subsection{SW Clump:  Mass Lower Limit Estimate}
We compute a lower limit estimate of the mass of the SW Clump by multiplying the mass of a spherical dust grain of average radius $\overline{a}$ with a lower limit number $N$ of grains in the Clump.  The number $N$ of grains in the Clump is at least the ratio of the total flux of the Clump at \Ks ~to the flux of scattered light from a single spherical grain with effective cross-section $Q_{sc}(\mbox{\Ks}) \cdot \pi \overline{a}^2$.  For a power law distribution of grain sizes which produces the dotted curve in the bottom plot of Figure \ref{DR_stack}, the average grain radius is $\overline{a}$ = 0.28 $\micron$ and the scattering efficiency of the distribution is $Q_{sc}$(\Ks) = 0.64.   Assuming a gas-to-dust mass ratio of 100:1 (e.g., \citet{Knapp:1993}) and a typical silicate mass density of $\rho$ = 3 gm cm$^{-3}$, we find the Clump's mass to have a lower limit of $M \gtrsim$ $5 \times 10^{-3}$ $M_\sun$.  If we assume the 500:1 gas-to-dust mass ratio found by \citet{Decin:2006} for VY CMa, this lower limit increases to $2.5 \times 10^{-2}$ $M_\sun$.  We emphasize that these are only lower limits on the Clump's mass; since the Clump is optically thick, its total mass is higher.  Compared to VY CMa's normal mass loss rate of $\sim$ $10^{-4}$ $M_{\sun}$ yr$^{-1}$, such a large mass in a single distinct feature is strongly indicative of an event involving ejection from a localized region of the star.

\subsection{Other Non-Detected Features}
As noted above, the feature designated as the S Arc by H07 is not detected in the \Ks ~image.  This places an upper limit of $\sim$ 0.06 for its scattering optical depth at \Ks.  Portions of the feature designated Arc 2 are marginally visible above the noise level at \Ks.  Their brightness corresponds to an optical depth no greater than about 0.2.  These optical depths are consistent with J07's finding these portions of Arc 2 to be optically thin in the visible.  In future observations, longer integration times might be able to detect scattered light emission from these features.

\section{CONCLUSION \& FUTURE WORK}
The \emph{HST} images of VY CMa made by S01 and H07 found the SW Clump to be completely obscured at visible wavelengths, emerging only at 1 $\micron$.  If the SW Clump were the product of a bipolar outflow with a similar feature opposite it in the heavily obscured region to the northeast of the star, such a feature would be more likley to appear in LMIRCam's \Ks, \Lp ~or M filters.  The absence of any obvious feature in the northeast opposite the Clump is consistent with VY CMa exhibiting a history of localized mass ejections from active regions on its surface, regions which are not strongly aligned with a presumed axis or equator.   H07 argued that the many features of VY CMa's ejecta may be the result of localized activity on the star related to convection and magnetic fields.  The distinct shape of the SW Clump is suggestive of a short-lived, localized event and may be analogous to a coronal mass ejection (CME) from a single location on the Sun's surface.  A short-lived ejection event is consistent with the SW Clump appearing as a confined, coherent shape several hundred years after ejection.  Adaptive optics imaging at wavelengths longer than 5 micron such as with NOMIC (8 - 25 $\micron$ camera on the LBT, \citep{Hinz:2012}) would better constrain the SED of the SW Clump and confirm whether its brightness at longer wavelengths is due to scattered light alone.

\vspace{12pt}
We thank the LBT staff for their hard work and support of these observations.  LMIRCam is funded by the National Science Foundation under grant NSF AST-0704992.  We thank the anonymous referee for helpful comments and feedback which have improved this manuscript.  D.S. is supported by funding from the United States Air Force.

\bibliographystyle{apj}
\bibliography{ms.bbl}

\end{document}